\documentclass[floatfix,aps,showpacs,nofootinbib,superscriptaddress]{revtex4}

\usepackage{graphicx}
\usepackage{epsfig}
\usepackage{mhchem}

\usepackage{amsmath}
\usepackage{amsfonts}
\usepackage{amssymb}
\usepackage{bm}
\usepackage{mathtools}

\usepackage{hyperref}
\usepackage{dsfont}
\usepackage{lipsum}

\usepackage{tikz}
\usepackage{circuitikz}
\usetikzlibrary{arrows, shapes}
\usepackage{pgfplots}
\pgfplotsset{compat=1.14}

\newcommand{\figpath}[1]{./figures/#1}

\newcommand{\mean}[1]{\left\langle #1 \right\rangle}

\newcommand{\Li}{\operatorname{Li}}

\begin{document}
 \author{Nahuel Freitas}
 \affiliation{Complex Systems and Statistical Mechanics, Department of Physics and Materials Science,
 University of Luxembourg, L-1511 Luxembourg, Luxembourg}
 \author{Gianmaria Falasco}
 \affiliation{Complex Systems and Statistical Mechanics, Department of Physics and Materials Science,
 University of Luxembourg, L-1511 Luxembourg, Luxembourg}
 \author{Massimiliano Esposito}
 \affiliation{Complex Systems and Statistical Mechanics, Department of Physics and Materials Science,
 University of Luxembourg, L-1511 Luxembourg, Luxembourg}

\title{Linear response in large deviations theory: \\
A method to compute non-equilibrium distributions}

\date{\today}

\begin{abstract}
We consider thermodynamically consistent autonomous Markov jump processes displaying a macroscopic limit in which the logarithm of the probability distribution is proportional to a scale-independent rate function (i.e., a large deviations principle is satisfied). In order to provide an explicit expression for the probability distribution valid away from equilibrium, we propose a linear response theory performed at the level of the rate function. We show that the first order non-equilibrium contribution to the steady state rate function, $g(\bm{x})$, satisfies $\bm{u}(\bm{x})\cdot \nabla g(\bm{x}) = -\beta \dot W(\bm{x})$ where the vector field $\bm{u}(\bm{x})$ defines the macroscopic deterministic dynamics, and the scalar field $\dot W(\bm{x})$ equals the rate at which work is performed on the system in a given state $\bm{x}$. 
This equation provides a practical way to determine $g(\bm{x})$, significantly outperforms standard linear response theory applied at the level of the probability distribution, and approximates the rate function surprisingly well in some far-from-equilibrium conditions. The method applies to a wealth of physical and chemical systems, that we exemplify by two analytically tractable models -- an electrical circuit and an autocatalytic chemical reaction network -- both undergoing a non-equilibrium transition from a monostable phase to a bistable phase.
Our approach can be easily generalized to transient probabilities and non-autonomous dynamics. Moreover, its recursive application generates a virtual flow in the probability space which allows to determine the steady state rate function arbitrarily far from equilibrium.
\end{abstract}

\maketitle

\section{Introduction}

Equilibrium statistical mechanics provides the probability distribution over the states of a system in terms of their energy.
Finding an explicit form that generalizes the Gibbs distribution to non-equilibrium stationary conditions has been the object of extensive research in the past and remains a central issue in non-equilibrium physics nowadays \cite{lebowitz1957, lebowitz1959, mclennan1959, zubarev1974, zubarev1994, komatsu2008, maes2010, colangeli2011, dhar2012, ness2013, falasco2021}.
The efforts in this direction may be divided into two research lines.
The first is based on expanding the non-equilibrium probability around the known equilibrium distribution.
Close to equilibrium, the correction to the Gibbs distribution
was first related to the dissipation of the system by McLennan  \cite{mclennan1959}. Since the dissipation is an extensive quantity of the trajectory duration and is quadratic in the distance from equilibrium, more recent work \cite{maes2010} has been devoted to clarify which part of the dissipation enter the McLennan formula and the suitable ordering of the long time and near to equilibrium limits.
Far from equilibrium, the stationary distribution has been obtained as formal series either based on all the cumulants of the (transient) entropy production or involving the excess dynamical activity induced by the non-equilibrium forces \cite{colangeli2011}.
Even at linear order around equilibrium, the difficulty in obtaining explicit results lies in the calculations of conditional averages over the full set of possible trajectories \cite{crooks2000, komatsu2008}. The second research line is based on
Freidlin-Wentzell large deviations theory \cite{ventsel1970},
which states that the logarithm of the stationary distribution of
systems with weak noise is proportional to a rate function, called
the quasi-potential. Since explicit computations of the rate function
are in general very difficult,  expansions around a reference
quasi-potential have been developed \cite{bouchet2016pert, tel1989}. However,
mainly systems affected by Gaussian noise were considered \cite{tel1989,bouchet2016pert}, and
without leveraging the thermodynamic structure of the underlying
dynamics.

In this paper we combine the previous two approaches by considering thermodynamically consistent stochastic dynamics displaying a macroscopic limit in which the probability distribution satisfies a large deviations principle.
Then, the steady state probability of a state $\bm{x}$ can be approximated by
$P_\text{ss}(\bm{x}) \propto \exp(-\Omega f(\bm{x}))$, for a sufficiently large scale parameter $\Omega$, where $f(\bm{x})$ is
the rate function or quasi-potential associated to the steady state. For a system in equilibrium at inverse temperature $\beta$, the usual Gibbs distribution indicates that $f(\bm{x})$ must be the free energy density
$\phi(\bm{x})$ times $\beta$ (assuming the free energy is extensive in $\Omega$).
Our first result is an explicit expression to compute the first order non-equilibrium correction $g(\bm{x})$ to the steady state rate function. We do so for systems subjected to Poissonian noise, i.e. microscopically described by a Markov jump process, thus complementing the large deviations approach and providing a clear
thermodynamic interpretation of the results.
We show that $g(\bm{x})$ is fully determined by the deterministic dynamics, described by a drift vector field
$\bm{u}(\bm{x})$, and the scalar field $\dot W(\bm{x})$ that indicates what is the rate at which work is performed on the
system for a given state $\bm{x}$.
This result offers a practical method to determine $g(\bm{x})$, which is applied to exactly solvable problems in electronics and chemistry , and is shown to be more accurate than usual linear response at the level of the probability distribution \cite{proesmans2015, proesmans2016, proesmans2019, brandner2015, bauer2016, tome2015}. It can also be generalized to transient evolutions and non-autonomous settings. Second, we extend our formalism and derive a general linear response formula for the quasi-potential around an arbitrary reference state.
Finally, we exploit these results to devise a virtual dynamics of the quasi-potential in a generic parameter space that, upon integration, allows one to calculate non-equilibrium rate functions starting from the known system free energy. This method considerably simplifies the complexity of the original problem, which amounts to the solution of an infinite-order algebraic equation in the gradient of the quasi-potential.

\section{Macroscopic limit and large deviations principle}

We consider a Markovian jump process on a discrete space $\bm{n} \in \mathbb{N}^k$ with time-independent transition rates $\lambda_\rho(\bm{n})$, associated to jumps $\bm{n} \to \bm{n} + \bm{\Delta}_\rho$. Namely, the time evolution of the probability distribution 
$P(\bm{n},t)$ over states is given by the master equation
\begin{equation}
\partial_t P(\bm{n}, t) = \sum_\rho \left[
\lambda_\rho(\bm{n} - \bm{\Delta}_\rho)
P(\bm{n} - \bm{\Delta}_\rho, t)
-\lambda_\rho(\bm{n}) P(\bm{n}, t) \right].
\label{apeq:master_eq}
\end{equation}
We assume that a macroscopic limit exists: there is a scaling parameter $\Omega$ such that
the transition rates scale as $\Omega$ and that the typical values of the density $\bm{x} = \bm{n}/\Omega$ are finite for $\Omega \to \infty$. Important examples satisfying the previous assumptions include chemical reaction networks  \cite{schmiedl2007, polettini2014, rao2016, lazarescu2019, avanzini2020, avanzini2021} in the large volume limit,
electronic circuits \cite{freitas2020, freitas2021} in the limit of large capacitances (see example in section \ref{sec:examples}-A), and some coarse-grained models of interacting many-body systems \cite{herpich2020}.
Then, in the limit $\Omega \to \infty$,
the probability distribution $P(\bm{x}, t)$ satisfies a large deviations (LD) principle \cite{touchette2009, Touchette2011}:
\begin{equation}
P(\bm{x}, t) \asymp e^{-\Omega f(\bm{x},t)}.
\label{eq:LD_principle}
\end{equation}
Indeed, plugging the previous ansatz in the master equation and keeping only the dominant terms
in $\Omega\to\infty$, we obtain the following dynamical equation for the rate function $f(\bm{x}, t)$ (see Appendix \ref{ap:derivation} for details):
\begin{equation}
\partial_t f(\bm{x}, t) = \sum_\rho \omega_\rho(\bm{x})
\left[1 - e^{\bm{\Delta}_\rho \cdot \nabla f(\bm{x}, t)}\right],
\label{eq:dyn_rate_func}
\end{equation}
where $\omega_\rho(\bm{x}) = \lim_{\Omega\to\infty} \lambda_\rho(\Omega \bm{x})/\Omega$ are the
scaled rates and the gradient operator is with respect to $\bm{x}$.
Thus, the steady state rate function $f_\text{ss}(\bm{x})$ must satisfy \cite{ge2016}:
\begin{equation}
0 = \sum_\rho \omega_\rho(\bm{x}) \left[1 - e^{\bm{\Delta}_\rho \cdot \nabla f_\text{ss}(\bm{x})}\right].
\label{eq:ss_rate_func}
\end{equation}
Eq. \eqref{eq:dyn_rate_func}
is a Hamilton-Jacobi equation for the action $f(\bm{x},t)$, with the Hamiltonian $H(\bm{x},\bm{p})=\sum_\rho \omega_\rho(\bm{x})
\left[1 - e^{\bm{\Delta}_\rho \cdot  \bm{p}}\right]$ and generalized momenta defined as usual as $\bm{p}=\nabla f$.
Therefore, the solutions of Eq. \eqref{eq:dyn_rate_func}
can also be found by integrating the associated Hamiltonian equations
\cite{graham1986, kamenev2011, cossetto2020}.
Eq. \eqref{eq:ss_rate_func} represents the stationary version, corresponding to a Hamiltonian dynamics on the manifold of constant null energy $H(\bm{x},\bm{p})=0$.
The steady state rate function solving the latter equation
is also known as a `quasi-potential' or `non-equilibrium potential'. This is
in part because of the analogy between Eq.~\eqref{eq:LD_principle} and the usual equilibrium Gibbs distribution, but more importantly because $f_\text{ss}(\bm{x})$ can be shown to be a Lyapunov function of the deterministic dynamics \cite{gang1986} (see Appendix \ref{ap:deterministic}), and provides information about the lifetime of non-equilibrium metastable states, in full analogy to equilibrium reaction-rate theory \cite{hanggi1990, bouchet2016,  cossetto2020, freitas2021}.

\section{Thermodynamic structure}

Thermodynamic consistency is enforced by assuming that for each transition with
rate $\lambda_\rho(\bm{n})$ there is a corresponding reverse transition with rate $\lambda_{-\rho}(\bm{n})$ (thus, $\bm{\Delta}_{-\rho} = -\bm{\Delta}_{\rho} $),
and that those rates are related by the local detailed balance (LDB) conditions
\begin{equation}
\log \frac{\lambda_\rho(\bm{n})}{\lambda_{-\rho}(\bm{n}+\bm{\Delta}_\rho)}
=
- \beta \left[\Phi(\bm{n} + \bm{\Delta}_\rho) - \Phi(\bm{n}) - W_\rho(\bm{n})\right],
\label{eq:ldb}
\end{equation}
where $\Phi(\bm{n})$
is the energy associated to a given state $\bm{n}$, and $W_\rho(\bm{n})$ is a non-conservative
work contribution associated to the transition $\bm{n} \to \bm{n} + \bm{\Delta}_\rho$. For simplicity we consider isothermal settings at inverse temperature $\beta$, but all of the following results hold also in more general situations where the system might interact with reservoirs at different temperatures. In that case, the LDB conditions take a form equivalent to Eq. \eqref{eq:ldb}, with the
energy $\beta\Phi(\bm{n})$ being replaced by a generalized Massieu potential, and the work contributions can be expressed in terms of a minimal set of fundamental forces \cite{rao2018}.

In the macroscopic limit we will assume that the energy of the system is an extensive quantity and therefore the limit
$\phi(\bm{x}) = \lim_{\Omega \to \infty}
\Phi(\bm{n} = \Omega \bm{x})/\Omega$ is well defined. For the work contributions we will abuse notation and write $W_\rho(\bm{x}) = \lim_{\Omega \to \infty} W_\rho(\bm{n} = \Omega \bm{x})$.
Thus, we consider situations in which
the work contributions associated to a transition do not
scale with the system size, as is indeed the case in
electronic circuits \cite{freitas2020, freitas2021}
and chemical reaction networks \cite{schmiedl2007, polettini2014, rao2016, lazarescu2019, avanzini2020, avanzini2021}.
According to the previous definitions, for large $\Omega$ the LDB conditions become:
\begin{equation}
\log \frac{\omega_\rho(\bm{x})}{\omega_{-\rho}(\bm{x})}
=
-\beta \left[\bm{\Delta}_\rho\cdot \nabla \phi(\bm{x}) - W_\rho(\bm{x})\right].
\label{eq:ldb_ld}
\end{equation}

\section{Deterministic dynamics}

From Eq.~\eqref{eq:LD_principle} we see that as we increase the scale parameter $\Omega$, the distribution $P(\bm{x},t)$ becomes increasingly
localized around the most probable states, i.e., the mimina of the rate function $f(\bm{x},t)$. If we let $\bm{x}(t)$ be one of the minima of $f(\cdot, t)$ for a given time $t$, then it is possible to show that it evolves according to the following closed deterministic dynamics:
\begin{equation}
    d_t \bm{x}(t) = \bm{u}(\bm{x}(t)),
    \label{eq:deterministic}
\end{equation}
where the deterministic drift $\bm{u}(\bm{x})$ is given by
\begin{equation}
    \bm{u}(\bm{x}) = \sum_\rho \omega_\rho(\bm{x}) \: \bm{\Delta}_\rho
    = \sum_{\rho>0} I_\rho(\bm{x}) \: \bm{\Delta}_\rho.
    \label{eq:drift}
\end{equation}
In the last equality we have exploited the fact that, due to the requirement of thermodynamic consistency, every transition has a reversed associated one, and expressed the drift $\bm{u}(\bm{x})$
in terms of the average forward currents $I_\rho(\bm{x}) = \omega_\rho(\bm{x}) - \omega_{-\rho}(\bm{x})$. Indeed,
a second order truncation in the Kramers-Moyal expansion of the master equation in Eq. \eqref{apeq:master_eq} leads to a Fokker-Planck equation with the vector field $\bm{u}(\bm{x})$ defined above as the drift. However, note that such expansion is not globally valid, but only locally around the minima of $f(\cdot,t)$ \cite{gardiner1985}.
An alternative proof of the previous result, based solely on Eq. \eqref{eq:dyn_rate_func}, is given in Appendix \ref{ap:deterministic}, where we also show that the steady state rate function satisfying Eq. \eqref{eq:ss_rate_func} is a Lyapunov function for the deterministic dynamics \cite{gang1986}.

\section{Equilibrium state and linear response}
\label{sec:equilibrium}

In this and the following section we will consider the perturbation of Eq. \eqref{eq:ss_rate_func} around the equilibrium steady state. For this, it is useful to define $\omega_\rho^{(0)}(\bm{x})$ to be the scaled rates in the absence of non-conservative sources of work (i.e., when $W_\rho(\bm{x}) =0$). According to Eq.~\eqref{eq:ldb_ld}, they satisfy
\begin{equation}
\log \frac{\omega^{(0)}_\rho(\bm{x})}{\omega^{(0)}_{-\rho}(\bm{x})}
=
-\beta \bm{\Delta}_\rho\cdot \nabla \phi(\bm{x}).
\label{eq:gdb_ld}
\end{equation}
In that case, the dynamics of the system is said to be (unconditionally) detailed balanced, and for long times a thermal equilibrium state
will be reached, which is exactly given by the Gibbs distribution
$P_\text{eq}(\bm{n}) \propto \exp(-\beta \Phi(\bm{n}))$. Thus, by Eq.~\eqref{eq:LD_principle},
the stationary rate function corresponding to this situation is $f_\text{ss}(\bm{x}) = \beta \phi(\bm{x})$.
In a general situation with non-vanishing work contributions $W_\rho(\bm{x})$, we can always consider the decomposition
\begin{equation}
f_\text{ss}(\bm{x}) = \beta \phi(\bm{x}) + g(\bm{x}),
\label{eq:split_f}
\end{equation}
where $g(\bm{x})$ quantifies the deviations with respect to thermal equilibrium. We will show in what follows that, to first order in the quantities $\beta W_\rho(\bm{x})$,  $g(\bm{x})$ must satisfy the equation
\begin{equation}
    \bm{u}^{(0)}(\bm{x})\cdot \nabla g(\bm{x}) = -\beta \dot W^{(0)}(\bm{x}),
    \label{eq:diff_eq_g}
\end{equation}
where the vector field $\bm{u}^{(0)}(\bm{x})$ and the scalar field $\dot W^{(0)}(\bm{x})$ are given by
\begin{equation}
    \bm{u}^{(0)}(\bm{x}) = \sum_{\rho>0} I^{(0)}_\rho(\bm{x}) \bm{\Delta}_\rho
    \qquad\qquad
    \dot W^{(0)}(\bm{x}) = \sum_{\rho>0} I^{(0)}_\rho(\bm{x}) W_\rho(\bm{x}),
\end{equation}
in terms of the currents
\begin{equation}
    I^{(0)}_\rho(\bm{x}) = \omega^{(0)}_\rho(\bm{x}) - \omega^{(0)}_{-\rho}(\bm{x}).
\end{equation}
Before proceeding with the proof of Eq. \eqref{eq:diff_eq_g}, we discuss the interpretation of the previous expressions. In the first place, $I^{(0)}_\rho(\bm{x})$ is just the average
current along transition $\rho$ according to
the detailed balanced dynamics, given that the state of the system is $\bm{x}$. Then,
the vector field $\bm{u}^{(0)}(\bm{x})$ is the net drift in state space for state $\bm{x}$, as discussed in the previous section.
As a consequence,
the field lines associated with $\bm{u}^{(0)}(\bm{x})$ are just the deterministic trajectories
of the detailed balanced system. The quantity $I^{(0)}_\rho(\bm{x}) W_\rho(\bm{x})$ is the
average work rate associated to the transition $\rho$ in the state $\bm{x}$ (to lower order in $W_\rho(\bm{x})$), and
$\dot W^{(0)}(\bm{x})$ is thus the total work rate.

Equation \eqref{eq:diff_eq_g} displays an apparent `gauge invariance', meaning that if $g(x)$ is any given solution, then ${\tilde g(\bm{x}) = g(\bm{x}) + h(\bm{x})}$ will also be a solution provided that $\bm{u}^{(0)}(\bm{x})\cdot \nabla h(\bm{x}) = 0$. However, this indeterminacy is removed if one demands that the function $g(\bm{x})$ must be a single-valued function for all $\bm{x}$. This can be seen by considering a situation in which the deterministic dynamics has a unique fixed point $\bm{x}^*$ (i.e., $\bm{u}(\bm{x}^*)=0$). Then, for any initial condition $\bm{x}_0$, the ensuing trajectory $\bm{x}(t|\bm{x}_0)$ satisfies $\lim_{t\to+\infty} \bm{x}(t|\bm{x}_0) = \bm{x}^*$. Also,
\begin{equation}
    d_t g(\bm{x}(t|\bm{x}_0)) =
    d_t \bm{x}(t|\bm{x}_0) \cdot \nabla g(\bm{x}(t|\bm{x}_0)) =
    \bm{u}^{(0)}(\bm{x}(t|\bm{x}_0)) \cdot \nabla g(\bm{x}(t|\bm{x}_0)) =
    -\beta \dot W^{(0)}(\bm{x}(t|\bm{x}_0)),
\end{equation}
which gives, once integrated over time,
\begin{equation}
    g(\bm{x}_0) = g(\bm{x}^*)  + \beta \int_0^{+\infty} \! \! dt  \: \dot W^{(0)}(\bm{x}(t|\bm{x}_0)).
\end{equation}
Thus, fixing the value $g(\bm{x}^*)$ at the fixed point, the previous equation allows to determine $g(\bm{x}_0)$ for any other point $\bm{x}_0$. This can be generalized to situations with multistability at equilibrium, where the deterministic dynamics has multiple fixed points. In that case one can construct local solutions $g(\bm{x})$
for each basin of attraction in the same way as before, fixing an arbitrary value for $g(\bm{x})$ at each fixed point. These values must be later adjusted in order for the different local solutions to match continuously with each other at the separatrices between different basins of attraction (the existence of a continuous quasi-potential for diffusive systems with multiple coexisting
attractors is dicussed in \cite{graham1986}).

\section{Proof}

We now proceed with the proof of Eq. \eqref{eq:diff_eq_g}. Since we are assuming thermodynamic consistency, we can rewrite Eq. \eqref{eq:ss_rate_func} as:
\begin{equation}
\begin{split}
0 &= \sum_{\rho>0}
\left\{
\omega_\rho(\bm{x}) \left[1 - e^{\bm{\Delta}_\rho \cdot \nabla f_\text{ss}(\bm{x})}\right] +
\omega_{-\rho}(\bm{x}) \left[1 - e^{-\bm{\Delta}_\rho \cdot \nabla f_\text{ss}(\bm{x})}\right]
\right\}\\
&=\sum_{\rho>0}
\left[\omega_\rho(\bm{x}) - \omega_{-\rho}(\bm{x}) e^{-\bm{\Delta}_\rho \cdot \nabla f_\text{ss}(\bm{x})} \right]
\left[1 - e^{\bm{\Delta}_\rho \cdot \nabla f_\text{ss}(\bm{x})}\right].
\end{split}
\end{equation}
Now we use the LDB conditions in Eq. \eqref{eq:ldb_ld} to write:
\begin{equation}
\begin{split}
0 &= \sum_{\rho>0}
\omega_\rho(\bm{x}) \left[1 - e^{\beta\left[\bm{\Delta}_\rho\cdot \nabla \phi(\bm{x}) - W_\rho(\bm{x})\right]} e^{-\bm{\Delta}_\rho \cdot \nabla f_\text{ss}(\bm{x})} \right]
\left[1 - e^{\bm{\Delta}_\rho \cdot \nabla f_\text{ss}(\bm{x})}\right].
\end{split}
\label{eq:proof_first_step}
\end{equation}
We see from the last expression that when $W_\rho(\bm{x})=0$,
the choice $f_\text{ss}(\bm{x}) = \beta \phi(\bm{x})$ is indeed a solution,
and in fact it makes each of the terms in the sum vanish independently. Thus, considering
$f_\text{ss}(\bm{x}) = \beta \phi(\bm{x}) + g(\bm{x})$, we have
\begin{equation}
0 = \sum_{\rho>0}
\omega_\rho(\bm{x}) \left[1 - e^{-\beta W_\rho(\bm{x})} e^{-\bm{\Delta}_\rho \cdot \nabla g(\bm{x})} \right]
\left[1 - e^{\beta \bm{\Delta}_\rho \cdot \nabla \phi(\bm{x})}
e^{\bm{\Delta}_\rho \cdot \nabla g(\bm{x})}\right].
\label{eq:proof_mid_step}
\end{equation}
Close to equilibrium $W_\rho(\bm{x})$ is small by definition
and $g(\bm{x})$ is assumed to be small and of the same order. It is therefore justified to expand Eq. \eqref{eq:proof_mid_step} to first order in both of them.
Doing that, we notice that the first factor in square brackets
becomes linear in $W_\rho(\bm{x})$ and $g(\bm{x})$, and therefore, to first order, all the other factors can be evaluated at zero order,
\begin{equation}
\begin{split}
0 &= \sum_{\rho>0}
\omega^{(0)}_\rho(\bm{x}) \left[\beta W_\rho(\bm{x})+\bm{\Delta}_\rho \cdot \nabla g(\bm{x})\right]
\left[1 - e^{\beta \bm{\Delta}_\rho \cdot \nabla \phi(\bm{x})}\right]\\
&= \sum_{\rho>0}
I^{(0)}_\rho(\bm{x}) \left[\beta W_\rho(\bm{x})+\bm{\Delta}_\rho \cdot \nabla g(\bm{x})\right],
\end{split}
\end{equation}
where in the last line we used the detailed balance conditions in Eq. \eqref{eq:gdb_ld} to
construct $I^{(0)}_\rho(\bm{x}) = \omega^{(0)}_\rho(\bm{x})-\omega^{(0)}_{-\rho}(\bm{x})$. The result of Eq. \eqref{eq:diff_eq_g} follows immediately.

\section{Dynamical extensions}
By considering a similar splitting of the time-dependent rate
function $f(\bm{x}, t) = \beta \phi(\bm{x}) + g(\bm{x}, t)$,
and repeating exactly the same derivation, this time starting from the dynamical equation of Eq. \eqref{eq:dyn_rate_func}, we obtain
the following dynamical equation for $g(\bm{x}, t)$:
\begin{equation}
    \partial_t g(\bm{x}, t) =
    \beta \dot W^{(0)}(\bm{x}) + \bm{u}^{(0)}(\bm{x})\cdot \nabla g(\bm{x}, t).
\end{equation}
We can also consider a situation with time dependent rates.
For a time-dependent scaled energy
$\phi(\bm{x}, t)$, $\beta \phi(\bm{x}, t)$ is the rate function that defines an instantaneous equilibrium state. Considering now the splitting
$f(\bm{x}, t) = \beta \phi(\bm{x},t) + g(\bm{x}, t)$,
we obtain
\begin{equation}
    \partial_t g(\bm{x}, t) =
    \beta [\dot W^{(0)}(\bm{x}, t) - \partial_t \phi(\bm{x}, t)]  + \bm{u}^{(0)}(\bm{x}, t)\cdot \nabla g(\bm{x}, t),
\end{equation}
where the time-dependent work rate $\dot W^{(0)}(\bm{x}, t)$
and drift $\bm{u}^{(0)}(\bm{x}, t)$ are defined in exactly the same way
as before, but in terms of the time-dependent currents
$I^{(0)}_\rho (\bm{x}, t) = \omega^{(0)}_\rho(\bm{x}, t) - \omega^{(0)}_{-\rho}(\bm{x}, t)$. Since the previous equation is a linear partial differential equation for $g(\bm{x}, t)$, for periodically driven systems one may also apply Floquet theory.

\section{Examples}
\label{sec:examples}

\subsection{CMOS SRAM cell}
\label{sec:bit}

\begin{figure*}
\centering
\begin{tikzpicture}
\node () at (0,0) {\includegraphics[width=\textwidth]{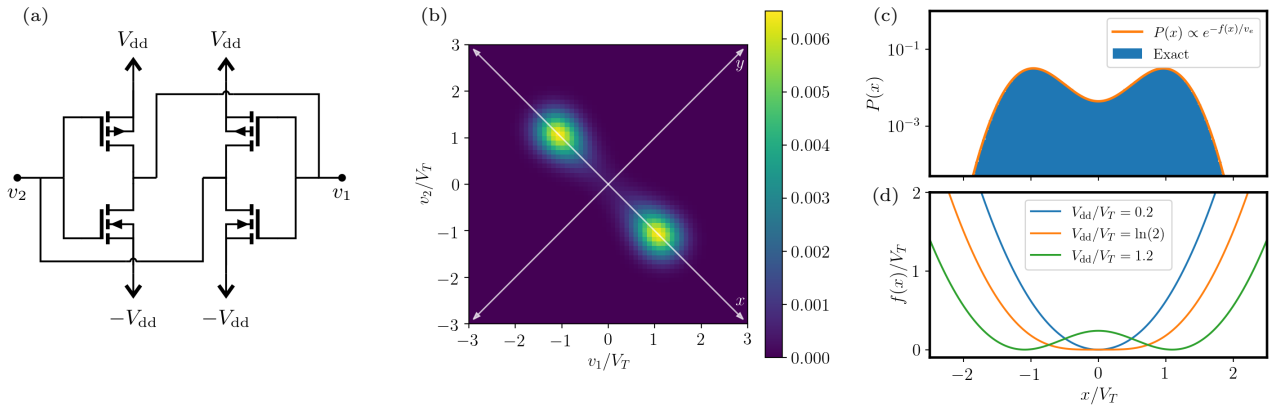}};
\node () at (-8,2.2) {\scriptsize (a)};
\node () at (-2.7, 2.2) {\scriptsize (b)};
\node () at (3.3, 2.2) {\scriptsize (c)};
\node () at (3.3, -.2) {\scriptsize (d)};
\end{tikzpicture}
\caption{
(a) Circuit diagram of a CMOS SRAM cell.
(b) 2D histogram of the steady state distribution
($V_\text{dd}/V_T=1.2$).
(c) Comparison of the exact steady state distribution for the variable
$x$ with the one obtained by the large deviations approximation
($V_\text{dd}/V_T=1.2$).
(d) Large deviations rate functions for different values of the powering voltage. In all cases the parameters are ${V_T = 26 \text{ mV}}$,  $v_e/V_T=0.1$, $\text{n}=1$.}
\label{fig:pbit_results}
\end{figure*}

We consider the model of a low-power static random access memory (SRAM) cell developed in \cite{freitas2020, freitas2021}. The usual CMOS implementation
of this kind of memories involves two inverters or NOT gates connected in a loop, each of which is composed of two MOS transistors, as shown in Figure \ref{fig:pbit_results}-(a). The memory is powered by applying a voltage bias $2V_\text{dd}$, which takes the system out of thermal equilibrium. The stochastic model of this device has two degrees of freedom: the charges $q_1$ and $q_2$ at the output of each inverter,
which correspond to voltages $v_{1/2} = q_{1/2}/C$, where $C$
is a value of capacitance characterizing the device.
Two Poisson rates are associated to each of the four transistors, corresponding to forward and backward conduction events.
In each conduction event, or jump, the voltages $v_{1/2}$ can change
by the elementary voltage $v_e = q_e/C$, where $q_e$ is the positive
electron charge.
Also, all Poisson rates are proportional to $I_0/q_e$, where
$I_0$ is a characteristic current. Under certain physical scalings of the transistors, as discussed in \cite{freitas2021}, both $I_0$ and $C$
increase linearly with the size of the device, and therefore we can take $C$, or equivalently $v_e^{-1}$, as the scaling parameter $\Omega$ above. Thus, the macroscopic limit $\Omega \to \infty$ must in this case be understood as the limit $v_e/V_T \to 0$ and $v_e/V_\text{dd} \to 0$,
where $V_T = (\beta q_e)^{-1}$ is the thermal voltage.
A typical steady state
distribution for this system (in its bistable phase, see below) is
shown in Figure \ref{fig:pbit_results}-(b). The rate function $f(v_1, v_2)$ corresponding to such steady state can be computed in terms of new variables $x=(v_1-v_2)/2$ and $y=(v_1+v_2)/2$. For example, the rate function of the reduced distribution for $x$ is exactly given by \cite{freitas2021}:
\begin{equation}
f(x) = \frac{x^2\!+\!2V_\text{dd}\:x }{V_T}
+ \frac{2\text{n} V_T}{\text{n}+2} \left[ L(x, V_\text{dd}) \!-\!L(x, -V_\text{dd})\right],
\label{eq:exact_f_bit}
\end{equation}
where $L(x,V_\text{dd}) = \Li_2\left(-\exp((V_\text{dd} + x(1+2/\text{n}))/V_T)\right)$, and $\Li_2(\cdot)$ is the polylogarithm function of second order, with $\text{n}$
a parameter characterizing the transistors.
In Figure \ref{fig:pbit_results}-(c) we compare the distribution obtained from the previous rate function with exact numerical results.
The agreement is essentially perfect even if only a few tens of electrons are involved (the scaling parameter is only $v_e=0.1V_T$,
and $V_\text{dd} = 1.15 V_T$).
Also, in Figure \ref{fig:pbit_results}-(c) we show that there is a
transition from a monostable phase into the bistable phase that allows the storage of a bit of information, occurring
at the critical powering voltage $V_\text{dd}^* = \ln(2) V_T$ for $\text{n}=1$.
Expanding the previous expression to first order in $V_\text{dd}$,
we obtain:
\begin{equation}
f(x)
= \underbrace{\frac{x^2}{V_T}}_{\beta \phi(x)}
\underbrace{
- 4 V_\text{dd} \frac{\text{n}}{\text{n}+2} \ln\left(2\cosh\left(\frac{\text{n}+2}{\text{2n}} \frac{x}{V_T}\right)\right)
+ \mathcal{O} (V_\text{dd}^2)}_{g(x)},
\label{eq:exp_f_bit}
\end{equation}
where we have identified the equilibrium and non-equilibrium contributions as in Eq. \eqref{eq:split_f}.  We will now recover this last result using the formalism developed here.

We begin by writing down the deterministic equations of motion, which read:
\begin{equation}
\begin{split}
    d_t v_1 &= I(v_1,v_2) - I(-v_1,-v_2) \\
    d_t v_2 &= I(v_2,v_1) - I(-v_2,-v_1), \\
\end{split}
\end{equation}
where the scaled current
$I(v_1, v_2)$
is defined in terms of the scaled rates $\omega_\pm(v_1,v_2)$
as $I(v_1, v_2) = \omega_+(v_1,v_2) - \omega_-(v_1,v_2)$. In turn,
these rates are given by:
\begin{equation}
\begin{split}
    \omega_+(v_1, v_2) &= (I_0/C) \:
    e^{(V_\text{dd} - v_2 - V_\text{th})/(\text{n}V_T)}\\
    \omega_-(v_1, v_2) &= \omega_+(v_1, v_2) \:
    e^{-(V_\text{dd} - v_1)/(V_T)}.
    \\
\end{split}
\end{equation}
Also, because of the symmetry of the device, the work contribution of each forward elementary transition is $q_e V_\text{dd}$ \cite{freitas2020},
and therefore the total work rate is
\begin{equation}
    \dot W(v_1, v_2) = q_e V_\text{dd}
    \left[I(v_1,v_2) + I(-v_1, -v_2) + I(v_2,v_1) + I(-v_2, -v_1)\right].
\end{equation}
To proceed, we write the equations of motion in the previously defined variables $x$ and $y$:
\begin{equation}
\begin{split}
    d_t x &= [I(x,y) - I(-x,y) - I(-x,-y) + I(x,-y)]/2\\
    d_t y &= [I(x,y) + I(-x,y) - I(-x,-y) - I(x,-y)]/2,\\
\end{split}
\end{equation}
where the change of variables in the function $I(\cdot)$ is implicit.
We see from the previous equations that if we consider an initial
condition where $y(0)=0$, then the ensuing trajectory will satisfy $y(t)=0$ at all times, since $d_t y|_{y=0} = 0$ for all $x$. This is true, in particular, for the equilibrium dynamics (that is, for $V_\text{dd}=0$)
which is the one we must consider in Eq. \eqref{eq:diff_eq_g}. Thus, over a trajectory with $y(0)=0$, Eq. \eqref{eq:diff_eq_g} reduces to:
\begin{equation}
d_t x^{(0)} \: d_x g(x) = -\beta \dot W^{(0)}(x),
\end{equation}
where the superscript (0) means in this case
that the currents should be evaluated
at $V_\text{dd}=0$. Then, we obtain:
\begin{equation}
\begin{split}
    g(x) &= -2 \frac{V_\text{dd}}{V_T} \int_0^x dx' \:
    \frac{I(x',0)+I(-x',0)}{I(x',0)-I(-x',0)}
    \bigg\rvert_{V_\text{dd}=0} \\
    &= - 4 V_\text{dd} \frac{\text{n}}{\text{n}+2} \ln\left(2\cosh\left(\frac{\text{n}+2}{\text{2n}} \frac{x}{V_T}\right)\right),
\end{split}
\label{eq:g_bit}
\end{equation}
in full agreement with Eq. \eqref{eq:exp_f_bit}.

\begin{figure*}
\centering
\includegraphics[width=\textwidth]{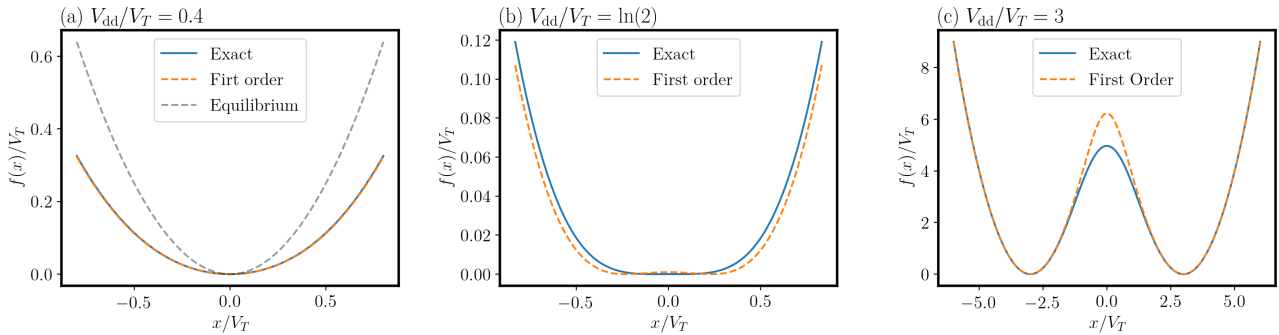}
\caption{Comparison for the CMOS SRAM cell between the exact rate function and the approximation to
first order in $V_\text{dd}/V_T$ for (a) $V_\text{dd}/V_T =0.4$ (monostable phase), (b) $V_\text{dd}/V_T = \ln(2)$ (critical point), and (c)
$V_\text{dd}/V_T = 3$ (bistable phase). In all cases, $\text{n}=1$.}
\label{fig:pbit_comparison}
\end{figure*}

In Figure \ref{fig:pbit_comparison} we compare the exact rate function of
Eq. \eqref{eq:exact_f_bit} with the first order approximation obtained with
our formalism. We see in Figure \ref{fig:pbit_comparison}-(a)
that while the approximation is a first order
expansion in $V_\text{dd}/V_T \ll 1$, it is essentially exact for
values as high as $V_\mathrm{dd}/V_T=0.4$. Even more remarkably, the first order approximation is able to predict the transition to bistability,
although as we see in Figure \ref{fig:pbit_comparison}-(b) the critical
point is underestimated (since at the exact critical point
$V_\text{dd}^*= \ln(2) V_T$ the approximation already developed a bistability). Finally, although the approximation is not expected
to work at all for $V_\text{dd}/V_T \gg 1$, we see
in Figure \ref{fig:pbit_comparison}-(c) that it correctly gets
the most probable values (which correspond to the fixed
points of the deterministic dynamics)
as well as the curvature around them, while it moderately fails to describe the height of the barrier separating the two fixed points.

\subsection{Schl\"ogl model}
\label{sec:schlogl}

We consider now the Schl\"ogl model \cite{vellela2009}, a well known autocatalytic  chemical model displaying bistability far from equilibrium.
It consists in the following set of chemical reactions,
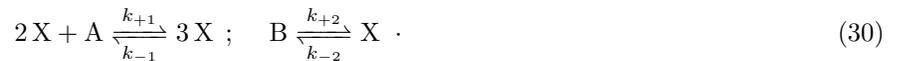
\begin{equation}
\ce{2X + A <=>[$k_{+1}$][$k_{-1}$] 3X}\:\: ; \: \quad \ce{ B  <=>[$k_{+2}$][$k_{-2}$] X \:.}
\end{equation}
 Here, species $A$ and $B$ will be considered to be chemostated reservoirs (i.e., we fix their concentration), and the only degree of freedom for the problem is the number of molecules of species $X$. The reactions are assumed to take place in a well-mixed container of volume $V$, which can be taken to be the macroscopic parameter $\Omega$. Thus, in the macroscopic limit we deal with the concentration $x$ of species $X$, and the scaled rates (ensuing the mass-action law) read:
\begin{align}
\omega_1(x)= k_{+1} a x^2, \quad  \omega_{-1}(x)= k_{-1} x^3,  \quad \omega_2(x)= k_{+2} b,  \quad \omega_{-2}(x)= k_{-2} x.
\end{align}
 We can set $k_{+1} a=1=k_{+2} b$ choosing appropriate units for time and concentrations. Then, in this case the LDB conditions impose the relation $k_{-1}=k_{-2}e^{-\Delta \mu}$, where
 $\Delta \mu$ is the chemical potential difference between species $A$ and $B$ (measured in units such that $\beta=1$ in the following). Thermal equilibrium is realized at $\Delta \mu=0$,
 for which the deterministic concentration of $x$ is $x^{(0)}=1/k_{-1}=1/k_{-2}$. The system can display bistability at $\Delta \mu \neq 0$ only for $k_{-2} \gtrsim 1.7 $ (so we will set $k_{-2}=2$ in the following plots).
 The exact rate function of the non-equilibrium steady state
 was obtained for example in \cite{vellela2009} and reads:
 \begin{align}
  f(x)&=\int_0^x \log \left(\frac{y k_{-2}(1+e^{-\Delta \mu} y^2)}{1+y^2}\right) \, dy  \\
  &=x \log \left(\frac{x k_{-2} (1+e^{-\Delta \mu } x^2)}{1+x^2}\right)+2 e^{\Delta \mu /2} \arctan \left(e^{-\frac{\Delta \mu}{2}} x\right)-x-2 \arctan(x)\\
  &=\underbrace{x\log(k_{-2}x)-x}_{\beta \phi(x)}
  \underbrace{+ \Delta \mu(\arctan(x)-x) + \mathcal{O}((\Delta \mu)^2)}_{g(x)}
  \label{eq:exact_rf_schlogl}
\end{align}
We now recover the previous result by employing our formalism.
The equilibrium currents are
\begin{align}
I^{(0)}_1(x)=(1-k_{-2}x)x^2, \quad I^{(0)}_2(x)=1-k_{-2}x
\end{align}
The work per reactions are $W_{1}=\Delta \mu$ and $W_{2}=0$, the total work rate is $\dot W^{(0)}(x)=\Delta \mu \: x^2(1-k_{-2}x)$ and $u^{(0)}(x)= (1-k_{-2}x)(1+x^2)$. Then, the equation for the first order correction to
the rate function is
\begin{align}
    (1-k_{-2}x)(1+x^2) \: d_x g(x) = -x^2(1-k_{-2}x)\Delta \mu,
\end{align}
from which we obtain
$g(x)= -\Delta \mu \int_0^x y^2/(1+y^2) \: dy = \Delta \mu [\arctan(x)-x]$,
in full agreement with Eq. \eqref{eq:exact_rf_schlogl}.

\begin{figure*}
\centering
\includegraphics[width=\textwidth]{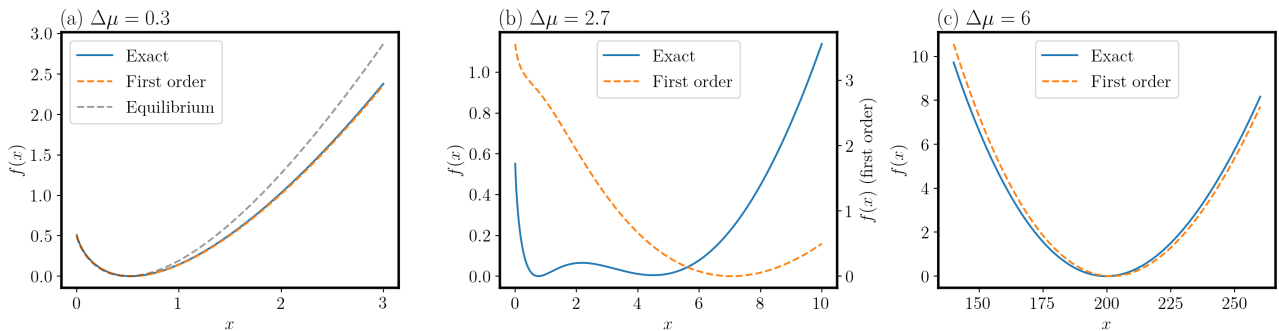}
\caption{Comparison between the exact rate function of the Schl\"ogl model and the approximation to first order in $\Delta \mu$. (a) For a monostable situation
close to equilibrium ($\Delta \mu = 0.3$). (b) For a bistable point
($\Delta \mu = 2.7$). (c) For large biases ($\Delta \mu = 6$). In all cases we
took $k_{-2} = 2$.}
\label{fig:schlogl_comparison}
\end{figure*}

In Figure \ref{fig:schlogl_comparison}-(a) we compare the equilibrium rate function with the exact and first order approximation for $\Delta \mu = 0.3$, showing that the first order approximation is indeed accurate. However, in contrast to the previous example, the first order approximation fails to offer
a even qualitatively description for values $\Delta \mu$ at which the exact rate functions develops a bistability (in fact, the first order approximation never develops a bistability). This model becomes monostable again for larger
values of $\Delta \mu$, and in that case the first order approximation correctly estimates the most probable value and the curvature around it, even if it is not expected to work in that regime, as was also observed in the previous example.

\section{Comparison with usual linear response}
\label{sec:comparison}

The approach that we presented exploits
the existence of a macroscopic limit with a LD principle and applies linear response to the LD rate function.
A natural question to ask is whether
this offers any advantage with respect to usual linear response theory \cite{proesmans2015, proesmans2016, proesmans2019, brandner2015, bauer2016, tome2015},
which directly gives the first order correction to the probability distribution (instead of the rate function associated to it), and does not require any macroscopic limit. We now show that our approach is more accurate than usual linear response and valid further away from equilibrium.

Assuming that the first order correction $g(\bm{x})$ is already known, the
LD approximation for the steady state distribution is:
\begin{equation}
\begin{split}
    P_\text{ss}(\bm{x}) & \simeq \frac{1}{Z} \: e^{-\Omega\beta f(\bm{x})}\\
    &\simeq \frac{1}{Z} \: e^{-\Omega\left(\beta \phi(\bm{x})+ g(\bm{x})\right)}\\
    & = \frac{1}{Z_\text{eq}} \: e^{-\Omega\beta \phi(\bm{x})}
    \: \frac{Z_\text{eq}}{Z} \: e^{ -\Omega g(\bm{x})}\\
    & \simeq P_\text{eq}(\bm{x}) \: \frac{Z_\text{eq}}{Z} \: e^{ -\Omega g(\bm{x})}.
\end{split}
\end{equation}
In the first line of the previous equation the LD approximation is involved, while in the second line the linear response (LR) approximation
at the level of the rate function is involved. Also,
$P_\text{eq}(\bm{x}) = \exp(-\Omega\beta\phi(\bm{x}))/Z_\text{eq}$ is the LD approximation of the equilibrium state. The partition function $Z$ is
\begin{equation}
    Z = \int d\bm{x}' \: e^{-\Omega \beta f(\bm{x}')} =
    \int d\bm{x}' \:  e^{-\Omega \beta \phi(\bm{x}')} e^{-\Omega g(\bm{x}') },
\end{equation}
and $Z_\text{eq} = Z|_{g(\bm{x})=0}$. For the LR approximation to be sensible we must have $g(\bm{x}) \ll \beta\phi(\bm{x})$.
However, if we consider the more stringent conditions $\Omega g(\bm{x}) \ll 1$, we can write
\begin{equation}
    P_\text{ss}(\bm{x}) \simeq P_\text{eq} (\bm{x})
    \left(
    1+
    \Omega \int d\bm{x}' \: P_\text{eq}(\bm{x}')  \left[g(\bm{x}') - g(\bm{x})\right]
    \right)
    +\mathcal{O} (\Omega^2 g^2),
\end{equation}
and
\begin{equation}
    Z \simeq Z_\text{eq}
    \left(
    1- \Omega \int d\bm{x}' \: P_\text{eq}(\bm{x}')  g(\bm{x}')
    \right)
    + \mathcal{O} (\Omega^2 g^2).
\end{equation}
From the previous equation we see that the first order perturbation to the probabilities, $\delta P(\bm{x}) = P_\text{ss}(\bm{x}) - P_\text{eq}(\bm{x})$, is related to the first order perturbation to the LD rate function by
\begin{equation}
    \delta P(\bm{x})/P_\text{eq}(\bm{x}) = -\Omega g(\bm{x}) + c,
    \label{eq:rel_dP_g}
\end{equation}
where $c$ is a constant ensuring $\int d\bm{x} \: \delta P(\bm{x}) = 0$.
Thus, we see that $\delta P(\bm{x})$ and $g(\bm{x})$ contain the same
information, as one can be computed from the other. However, in the
macroscopic limit ($\Omega \gg 1$), the LR approximation at the level of
the rate function is expected to be more accurate than the LR
approximation at the level of the probabilities, and also valid for
stronger perturbations. The reason is that for $\Omega \gg 1$ the
conditions $g(\bm{x}) \ll \beta\phi(\bm{x})$ might be satisfied even if
the conditions  $\Omega g(\bm{x}) \ll 1$ are not.

\begin{figure*}
\centering
\includegraphics[width=\textwidth]{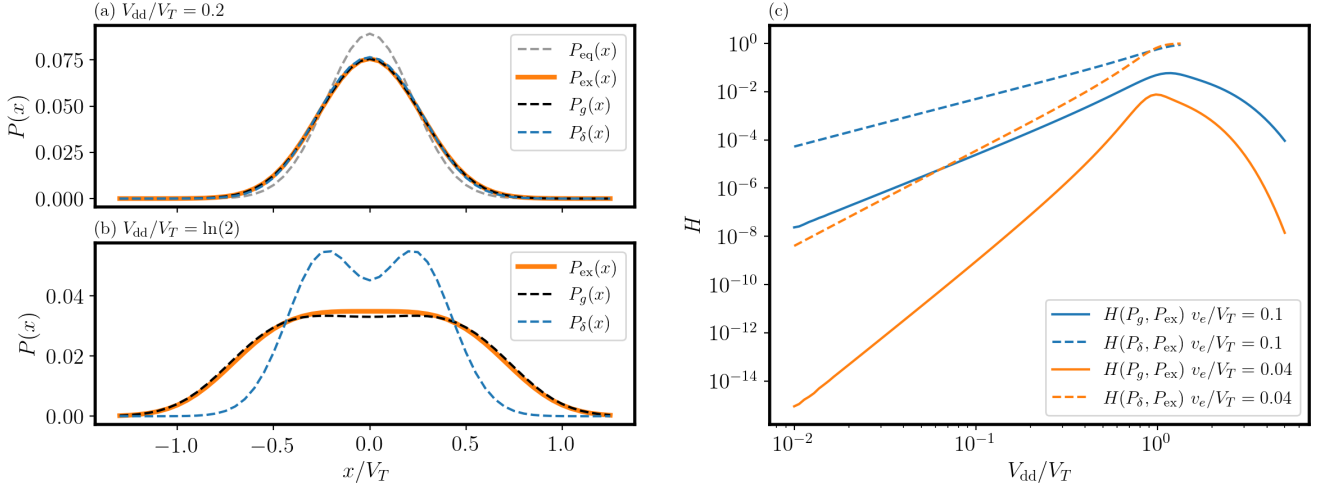}
\caption{(a) For the example of section \ref{sec:bit}, we compare the equilibrium distribution $P_\text{eq}(x)$ with the exact distribution
$P_\text{ex}(x)$, the LR approximation at the level of the rate function $P_g(x)$, and the LR approximation at the level of the probability distribution $P_\delta(x)$, for $V_\text{dd}/V_T=0.2$ and $v_e/V_T=0.1$. (b) Same as (a) but for $V_\text{dd}/V_T = \ln(2)$, which is the critical
point between the monostable and bistable phases. The equilibrium distribution is not shown in this case. (c) Hellinger distances
$H(P_g, P_\text{ex})$ and $H(P_\delta, P_\text{ex})$ as a function of
$V_\text{dd}$ for two different values of $v_e$.
}
\label{fig:usual_LR_comparison}
\end{figure*}

We now illustrate the previous results in the
example of section \ref{sec:bit} (CMOS SRAM cell). For a given value of the powering voltage $V_\text{dd}$, we will compare three probability distributions: i) The LD approximation with the exact rate function
of Eq. \eqref{eq:exact_f_bit}: $P_\text{ex}(x) \propto \exp(-f(x)/v_e)$ (note that, as shown in Figure \ref{fig:pbit_results}-(c), this distribution can be considered exact already for $v_e/V_T \simeq 0.1$), ii) The LD approximation combined with the LR approximation of the rate
function: $P_\text{g}(x) \propto \exp(-(x^2/V_T + g(x))/v_e)$, where
$g(x)$ is given by Eq. \eqref{eq:g_bit}, and iii)
the usual LR approximation at the level of the probability
distribution:
$P_\delta(x) = P_\text{eq}(x) + \delta P(x) =
P_\text{eq}(x) (1-g(x)/v_e + c)$ (see Eq. \eqref{eq:rel_dP_g}).
In Figure \ref{fig:usual_LR_comparison}-(a)
we see that for $V_\text{dd}/V_T = 0.2$, both $P_g(x)$ and $P_\delta(x)$
correctly describe the deviations from the equilibrium distribution.
However,
further away from equilibrium (in particular, at the
critical point $V_\text{dd} = V_\text{dd}^* = V_T\ln(2)$), $P_g(x)$
remains an acceptable approximation while $P_\delta(x)$ does not,
as shown in Figure \ref{fig:usual_LR_comparison}-(b).
Finally, in Figure \ref{fig:usual_LR_comparison}-(c) we compare
the Hellinger distances $H(P_g, P_\text{ex})$ and $H(P_\delta, P_\text{ex})$ between the exact distribution and the LR approximations at the level of the rate function or at the level of the probability distribution, respectively, as a function of the powering voltage
$V_\text{dd}$ (the Hellinger distance $0\leq H(p,q) \leq1$ between distributions $p(x)$ and $q(x)$ is defined by $H^2(p,q) = 1-\sum_i\sqrt{p(x_i)q(x_i)}$).
We see that $P_g(x)$ is always a better approximation with respect to $P_\delta(x)$, and that the gap between the two increases
as we go deeper into the macroscopic limit (i.e., if we decrease the value of the elementary voltage $v_e$), as predicted above. We note that for sufficiently large $V_\text{dd}$ the function $P_\delta(x)$ stops being
positive definite (a common issue of the usual LR approximation at the
level of the probability distribution) and therefore the Hellinger
distance cannot be computed. Also, in this particular case the
distance $H(P_g, P_\text{ex})$ reaches a maximum for large $V_\text{dd}$,
after which it decreases again. This is a consequence of the observation made in the previous section that the LR approximation at the level of the rate function correctly describes the most probable values and the curvature around them for large biases, even if a priori it is not expected to work in that regime.

\section{Linear response of a non-equilibrium steady state}

In this section we generalize our previous results by considering the
problem of perturbing a general non-equilibrium steady state of an autonomous system. The rate function $f(\bm{x})$ of the steady state corresponding to some given work contributions $W_\rho(\bm{x})$
must satisfy Eq. \eqref{eq:proof_first_step}:
\begin{equation}
\begin{split}
0 &= \sum_{\rho>0}
\omega_\rho(\bm{x}) \left[1 - e^{\beta\left[\bm{\Delta}_\rho\cdot \nabla \phi(\bm{x}) - W_\rho(\bm{x})\right]} e^{-\bm{\Delta}_\rho \cdot \nabla f(\bm{x})} \right]
\left[1 - e^{\bm{\Delta}_\rho \cdot \nabla f(\bm{x})}\right].
\end{split}
\end{equation}
Let us now consider the perturbation
$W_\rho(\bm{x}) \to W'_\rho(\bm{x}) = W_\rho(\bm{x}) + \delta W_\rho(\bm{x})$, and the corresponding response
${f(\bm{x}) \to f'(\bm{x}) = f(\bm{x}) + g(\bm{x})}$.
By expanding the previous equation to first order in
$\delta W_\rho(\bm{x})$ and $g(\bm{x})$, we find
\begin{equation}
\bm{\mathcal{U}}(\bm{x}) \cdot \nabla g(\bm{x}) = -\beta \mathcal{\dot W}(\bm{x}).
\label{eq:diff_eq_g_ss}
\end{equation}
The previous equation has the same structure as
Eq. \eqref{eq:diff_eq_g} for perturbations around equilibrium,
but this time the vector field $\bm{\mathcal{U}}(\bm{x})$
and the scalar field $\mathcal{\dot W}(\bm{x})$ are given by
\begin{equation}
\bm{\mathcal{U}}(\bm{x}) =
\sum_{\rho>0}
\left(\omega_{-\rho}(\bm{x}) e^{-\Delta_\rho \cdot \nabla f(\bm{x})} -
\omega_{\rho}(\bm{x}) e^{\Delta_\rho \cdot \nabla f(\bm{x})}\right)\bm{\Delta}_\rho,
\label{eq:def_U_ss}
\end{equation}
and
\begin{equation}
\mathcal{\dot W}(\bm{x}) =
\sum_{\rho>0} \omega_{-\rho}(\bm{x}) \left(e^{-\Delta_\rho \cdot \nabla f(\bm{x})} -1\right)
\delta W_\rho(\bm{x}).
\label{eq:def_W_ss}
\end{equation}
The previous results of section \ref{sec:equilibrium} are easily recovered by considering
$f(\bm{x}) \to \beta \phi(\bm{x})$, $W_\rho(\bm{x}) \to 0$,
and using the LDB conditions at equilibrium,
since in that way we see that
$\bm{\mathcal{U}}(\bm{x})$ reduces
to the equilibrium deterministic drift $\bm{u}(\bm{x})$, and
$\mathcal{\dot W}(\bm{x})$ to the work rate $\dot W(\bm{x})$. We note that $\bm{\mathcal{U}}(\bm{x})$ does not correspond to the field of deterministic trajectories, and also that $\mathcal{\dot W}(\bm{x})$
cannot be interpreted as the average work rate for a given state, as
is the case for perturbations of the equilibrium distribution.
However, it is interesting to note that
the fixed points of the true deterministic dynamics are shared
by the dynamics corresponding to the alternative drift
$\bm{\mathcal{U}}(\bm{x})$, and also that $\mathcal{\dot W}(\bm{x})$
vanishes at those fixed points (see below).

\section{Virtual evolution of non-equilibrium steady states}
\label{sec:evolution}

\begin{figure*}
\centering
\includegraphics[width=\textwidth]{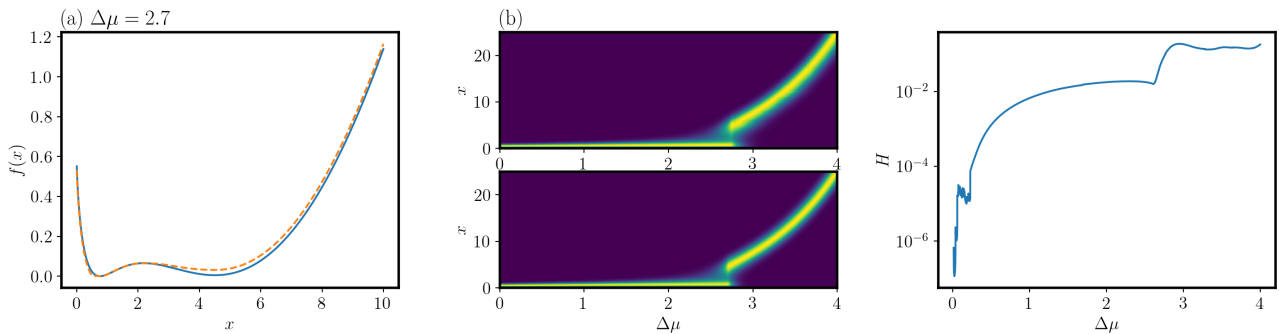}
\caption{(a) Comparison between the exact rate function of the Schl\"ogl model for $\Delta \mu = 2.7$ and the one obtained numerically with the method explained in section \ref{sec:evolution}. (b) In the top panel we show the full virtual evolution of the probability distribution up to $\Delta \mu =4$ and in the
bottom panel we show the corresponding exact results (see details in the text). (c) Hellinger distance between the steady state probability distributions obtained from the evolved and exact rate functions, as a function of $\Delta \mu$. In all cases we took $k_{-2}=2$.}
\label{fig:evolution}
\end{figure*}

In equilibrium statistical mechanics the unnormalized
thermal distribution $P(\bm{x} | \beta) = \exp(-\beta \Phi(\bm{x}))$
at a given inverse temperature $\beta$
can be obtained by solving the evolution equation
$d_\beta P(\bm{x} | \beta) = -\Phi(\bm{x}) P(\bm{x} | \beta)$,
with a known initial condition (typically, the infinite temperature
distribution $P(\bm{x} | 0) = 1$). Thus, the previous differential equation allows to obtain the distribution for an inverse temperature
$\beta + d \beta$ given the knowledge of the distribution at inverse
temperature $\beta$. Interestingly, Eq. \eqref{eq:diff_eq_g_ss} provides the basis for a similar procedure, in which a non-equilibrium steady state distribution is evolved starting, for example, from the equilibrium one.
In order to illustrate this idea we will consider a system with a single state-independent work parameter $\lambda$ (as the examples given in
section \ref{sec:examples}). Then, the steady state rate function
$f(\bm{x}|\lambda)$ has a parametric dependence on $\lambda$, and we can rewrite
Eq. \eqref{eq:diff_eq_g_ss} as:
\begin{equation}
    \bm{\mathcal{U}}(\bm{x}|\lambda) \cdot \nabla
    [\partial_\lambda f(\bm{x}|\lambda)] = -\beta \mathcal{\dot W}(\bm{x}|\lambda).
    \label{eq:diff_eq_evol}
\end{equation}
Solving this equation we can obtain $\partial_\lambda f(\bm{x}|\lambda)$
up to an irrelevant constant, and if $f(\bm{x}|\lambda)$ is known,
we can approximate
$f(\bm{x}|\lambda + d\lambda) \simeq f(\bm{x}|\lambda) +
d\lambda \: \partial_\lambda f(\bm{x}|\lambda)$. Iterating this procedure it is in principle possible to obtain the rate function
$f(\bm{x}|\lambda)$ for arbitrary $\lambda$ starting from the
equilibrium distribution $f(\bm{x}|0)$.

We now apply this method to the Schl\"ogl model discussed in section
\ref{sec:schlogl}. In that case the work parameter is
$\lambda = \Delta \mu$ and, since the problem is one-dimensional,
we can write Eq. \eqref{eq:diff_eq_evol} as:
\begin{equation}
    \partial_{\Delta \mu} f(x|\Delta \mu)
    = -\beta \int_0^x  \:
    \frac{\mathcal{\dot W}(x'|\Delta \mu)}
    {\mathcal{U}(x'|\Delta \mu) } \: dx'.
\end{equation}
We numerically solve the previous differential equation using
a Runge-Kutta method of order 4, starting from the equilibrium
rate function $f(x|0)=x\log(k_{-2} x)-x$ (see Eq. \eqref{eq:exact_rf_schlogl}). In Figure \ref{fig:evolution}-(a) we compare the exact rate function given
by Eq. \eqref{eq:exact_rf_schlogl} with the one obtained with our evolution method up to $\Delta \mu = 2.7$, a bistable point of the system. To obtain the evolved rate function shown in the plot
we have sampled the intervals $x\in [0,10]$ and
$\Delta \mu \in [0,2.7]$ at $4\times 10^3$ points each.
In the top panel of Figure \ref{fig:evolution}-(b) we
show the results obtained
by performing the evolution up to $\Delta \mu=4$,
this time sampling the interval $x\in [0,25]$ at $10^4$ points and
the interval $\Delta \mu \in [0,4]$ at $4\times 10^3$ points.
The color plot indicates the value of the probability distribution
$P_\text{ev}(x|\Delta \mu) \propto e^{-\Omega f(x|\Delta \mu)}$ (with $\Omega=10$), where $f(x|\Delta \mu)$ is the rate function obtained with our method at each point in the evolution. For visualization purposes, it was normalized so that the maximum of $P_\text{ev}(x|\Delta \mu)$ for a given $\Delta \mu$ is always $1$. In the bottom panel of Figure \ref{fig:evolution}-(b) we show the exact probability distribution $P_\text{ex}(x|\Delta \mu)$ computed in the same way as before but in terms of the exact rate function in Eq. \eqref{eq:exact_rf_schlogl}. In Figure \ref{fig:evolution}-(c)
we show the Hellinger distance $H(P_\text{ev}, P_\text{ex})$
as a function of $\Delta \mu$. We see that the distance between the distributions computed from the evolved and exact rate functions increases rapidly at the beginning of the evolution, reaches a plateau, and increases again after the bistable region around $\Delta \mu \simeq 2.7$ is crossed.

An important comment regarding the numerical stability of the proposed method is in order. For Eq. \eqref{eq:diff_eq_g_ss} to have a smooth and well behaved solution (that is, such that $\nabla g(\bm{x})$ is finite for all $\bm{x}$),
the field $\mathcal{\dot W}(\bm{x})$ must vanish wherever the
field $\bm{\mathcal{U}}(\bm{x})$ does. As already noted in the previous section, if $\bm{x}^*$ is a fixed point of the deterministic dynamics,
then from Eqs. \eqref{eq:def_U_ss} and \eqref{eq:def_W_ss} we see that $\bm{\mathcal{U}}(\bm{x}^*)=0$ and $\mathcal{\dot W}(\bm{x}^*)=0$.
This follows from the LDB conditions and the fact that $\nabla f(\bm{x}^*)=0$ (see Appendix \ref{ap:deterministic}). Thus,
assuming that the deterministic fixed points are the only zeros of
$\bm{\mathcal{U}}(\bm{x})$, the previous minimal consistency condition is
indeed satisfied by Eqs. \eqref{eq:def_U_ss} and \eqref{eq:def_W_ss}. Importantly,
for this to hold, the unperturbed rate function $f(\bm{x})$ involved in the definitions of $\bm{\mathcal{U}}(\bm{x})$ and $\mathcal{\dot W}(\bm{x})$
must correspond exactly to the parameters at which the rates $\omega_\rho(\bm{x})$ are evaluated. However, that correspondence is broken
in the intermediate steps of our iterative procedure, since at each step (except at the first one when the equilibrium distribution is perturbed) the fields
$\bm{\mathcal{U}}(\bm{x})$ and $\mathcal{\dot W}(\bm{x})$ are constructed from
an approximate rate function $f(\bm{x})$, which might not exactly satisfy $\nabla f(\bm{x}^*) = 0$. This issue might lead to errors and discontinuities for values
of $\bm{x}$ close to the deterministic fixed points at each step, which will propagate forward in the virtual evolution. That is in fact the main source of errors contributing to the deviations between the exact and evolved rate functions in Figure \ref{fig:evolution}-(a). A more sophisticated numerical implementation of the proposed method might take as an input the deterministic
fixed points $\{\bm{x}^*\}$ at each step, in order to ensure that they match the
minima of the approximated rate function. This point should be considered for applications to more complex systems in many dimensions.

\section{Conclusion}

For thermodynamically consistent stochastic systems displaying a macroscopic limit in which the stationary distribution fulfills a large deviations principle, we have
shown how to compute the correction to the rate function, or quasi-potential, to first order in the forces taking the system out of thermal equilibrium.
We have applied this result to two examples: a realistic model of an important kind of electronic memory and the Schl\"ogl model of a bistable chemical reaction. We also generalized our methods to time-dependent settings. In the final part of our work we considered the linear response of arbitrary non-equilibrium steady states. Based on that result, we developed a method to incrementally evolve the stationary state rate function in the space of thermodynamic forces, which in principle allows to obtain the rate function arbitrarily far from equilibrium, starting from the known equilibrium one. This might lead to new numerical schemes to study non-equilibrium fluctuations at the mesoscopic level, providing an alternative to the usual stochastic simulations based on the Gillespie algorithm.

Our work makes use of the thermodynamic consistency of the dynamics and of the existence of a macroscopic limit.
The results might remain nonetheless an excellent approximation even significantly away from the strict macroscopic limit, as is the case in the examples discussed in section \ref{sec:examples}.
Furthermore, we have shown that in the mesoscopic or macroscopic regime the linear response approach at the level of the rate function is more accurate than traditional linear response theory at the level of the probability distribution, even if both approaches are first order expansions in the thermodynamic forces. As shown in the example in 
section \ref{sec:comparison} (Figure \ref{fig:usual_LR_comparison}), what actually happens is that both approaches become more accurate as the scale of the system is increased, but with a larger gap between the accuracy of linear response at the level of the rate function and at the level of the probability distribution. 
This is understood as follows. A fixed change $\Delta \bm{x}$ in the density $\bm{x}=\bm{n}/\Omega$ requires an increasing number of elementary transitions as the scale of the system is increased. Also, the work contribution during a transition does not scale with the system size, while the associated change in free energy does. Thus, it is natural to expect linear response approximations to be more accurate in the macroscopic regime, since they are based on the assumption that work contributions are small with respect to other energy scales. However, while the first order non-equilibrium correction to the rate function is independent of the scale (by definition of the rate function), the first order correction to the probability distribution 
$P(\bm{x})$ is proportional to the scale parameter (see Eq. \eqref{eq:rel_dP_g}), and that constrains the strength of non-equilibrium forces for which the correction remains small.

\section{Acknowledgments}
We acknowledge funding from the INTER project  ``TheCirco'' (INTER/FNRS/20/15074473) and CORE project ``NTEC'' (C19/MS/13664907), funded by the Fonds National de la Recherche (FNR, Luxembourg), and from the European Research Council,
project NanoThermo (ERC-2015-CoGAgreement No. 681456).

\appendix

\section{Dynamics of the large deviations rate function}
\label{ap:derivation}
In this section we provide details of the derivation of Eq. \eqref{eq:dyn_rate_func} in the main text, starting from the master equation in Eq. \eqref{apeq:master_eq}. In the first place we represent a given discrete state $\bm{n}$ of the system by its `density' $\bm{x} = \bm{n}/\Omega$. For example, the components of $\bm{n}$ are the numbers of molecules in a chemical reaction network or the numbers of charges in an electronic circuit, while the components $\bm{x}$ are, respectively, the concentrations or the voltages. Thus, a transition $\bm{n} \to \bm{n} + \bm{\Delta}_\rho$
corresponds to a change $\bm{\Delta}_\rho/\Omega$ in the density $\bm{x}$.  We consider the large deviations 
ansatz $P(\bm{x},t) = \exp(-\Omega f(\bm{x},t))/Z(t)$ for the time-dependent probability distribution in terms of $\bm{x}$, where $Z(t) = \sum_{\bm{n}} \exp(-\Omega f(\bm{n}/\Omega,t))$ takes care of the normalization.
Then, we have 
\begin{equation}
P(\bm{x} + \bm{\Delta}_\rho/\Omega,t) = 
P(\bm{x},t)e^{-\bm{\Delta}_\rho\cdot \nabla f(\bm{x})
+\mathcal{O}(\Omega^{-1})},
\end{equation}
and
\begin{equation}
d_t P(\bm{x},t) = 
 -P(\bm{x},t) [\Omega \: d_t f(\bm{x},t)
  + d_t \log(Z(t))] .
\end{equation}
Eq. \eqref{eq:dyn_rate_func} 
is then obtained by 
inserting the previous two expressions in the master equation of Eq. \eqref{apeq:master_eq}, 
dividing both sides by $\Omega P(\bm{x},t)$,
and taking the limit $\Omega \to \infty$ in which 
$\Omega^{-1} \log(Z(t)) \to 0 $.

\section{Deterministic dynamics}
\label{ap:deterministic}

We define $\bm{x}(t)$ to be a minimum of the rate function $f(\cdot, t)$ at time
$t$. Then, we have that $\partial_{x_i} f(\bm{x}(t), t) = 0$ and
that the symmetric
matrix with elements $[\partial^2_{x_i, x_j} f(\bm{x}(t), t)]_{ij}$ is
positive semidefinite. For time $t + dt$ we can write:
\begin{equation}
\begin{split}
    0 &= \partial_{x_i} f(\bm{x}(t + dt), t+dt) \\
    &= \partial_{x_i} f(\bm{x}(t) + \bm{u}(t) \: dt+ \mathcal{O}(dt^2), t+dt) \\
    &= u_j(t) \: \partial^2_{x_j,x_i} f(\bm{x}(t),t) \: dt
    + \partial_t \partial_{x_i} f(\bm{x}(t),t) \: dt + \mathcal{O}(dt^2),
\end{split}
\label{ap:first_step}
\end{equation}
where we have introduced the velocity $\bm{u}(t) = d_t \bm{x}(t)$, and repeated indices are summed. Now we employ the differential equation satisfied by
$f(\bm{x},t)$, Eq. \eqref{eq:dyn_rate_func} in the main text, to obtain:
\begin{equation}
\begin{split}
    \partial_t \partial_{x_i} f(\bm{x},t) &=
    \partial_{x_i} \partial_t f(\bm{x},t) \\
    &= \partial_{x_i}
    \left[ \sum_\rho \omega_\rho(\bm{x}) \left[1-e^{(\bm{\Delta}_\rho)_j \partial_{x_j}f(\bm{x},t)}\right]\right]\\
    & = \sum_\rho \partial_{x_i} \omega_\rho(\bm{x})
    \left[1-e^{(\bm{\Delta}_\rho)_j \partial_{x_j}f(\bm{x},t)}\right]
    +
    \sum_\rho \omega_\rho(\bm{x})
    \left[-e^{(\bm{\Delta}_\rho)_j \partial_{x_j}f(\bm{x},t)}
    (\bm{\Delta}_\rho)_j \partial^2_{x_i,x_j}f(\bm{x},t)\right].
\end{split}
\end{equation}
Noticing that the first sum in the last line of the previous equation vanishes when evaluated at $\bm{x}(t)$, and combining the result with
Eq. \eqref{ap:first_step}, up to first order in $dt$ we obtain:
\begin{equation}
    \left[\bm{u}(t) - \sum_\rho \omega_\rho(\bm{x}(t)) \bm{\Delta}_\rho\right]_j
    \partial^2_{x_j,x_i}f(\bm{x}(t),t) = 0.
\end{equation}
Thus, whenever the matrix $[\partial^2_{x_i, x_j} f(\bm{x}(t), t)]_{ij}$
can be assumed to be positive definite, we have
$\bm{u}(t) - \sum_\rho \omega_\rho(\bm{x}(t)) \bm{\Delta}_\rho = 0$, in
agreement with Eqs. \eqref{eq:deterministic} and \eqref{eq:drift} in the main text.

If the rate function has a single global minimum, then the instantaneous average value
$\mean{\bm{x}(t)} = \Omega^{-1} \mean{\bm{n}(t)} = \Omega^{-1} \sum_{\bm{n}} \bm{n} P(\bm{n},t)$
also evolves as $d_t \mean{\bm{x}(t)} = \bm{u}(\mean{\bm{x}(t)})$ in the
$\Omega \to \infty$ limit. In order to see this, we multiply
Eq. \eqref{apeq:master_eq} in the main text by $\bm{n}$ and sum, obtaining:
\begin{align}\label{eq:mean}
 \partial_t \sum_{\bm{n}}  \bm{n} P(\bm{n}, t) &= \sum_{\rho,\bm{n}} \bm{n} \left[
\lambda_\rho(\bm{n} - \bm{\Delta}_\rho)
P(\bm{n} - \bm{\Delta}_\rho, t)
-\lambda_\rho(\bm{n}) P(\bm{n}, t) \right]\\
&= \sum_{\rho,\bm{n}}\left[( \bm{n} +\bm{\Delta}_\rho )
\lambda_\rho(\bm{n} )
P(\bm{n} , t)
-\bm{n}\lambda_\rho(\bm{n}) P(\bm{n}, t) \right]\\
&=\sum_{\rho,\bm{n}}  \bm{\Delta}_\rho
\lambda_\rho(\bm{n} )
P(\bm{n} , t),
\end{align}
where we shifted $\bm{n} \to \bm{n} +\bm{\Delta}_\rho$ in the first sum on the RHS of \eqref{eq:mean}.
Dividing both sides by $\Omega$, passing to integral over $\bm{x}$ and recognizing that $P$ peaks around the single minimum of $f$ as  $\Omega \to \infty$ we obtain the desired result.

Finally, we mention that if $f_\text{ss}(\bm{x})$ is the steady state rate function, i.e., the solution to Eq. \eqref{eq:ss_rate_func}, then it follows that \cite{gang1986}
\begin{equation}
    d_t f_\text{ss}(\bm{x}(t))
    = \sum_\rho \omega_\rho(\bm{x}) \:
    \bm{\Delta}_\rho\cdot\nabla f_\text{ss}(\bm{x}(t)) 
    \leq \sum_\rho  \omega_\rho(\bm{x}(t))
    [e^{\bm{\Delta}_\rho\cdot f_\text{ss}(\bm{x})}-1] = 0,
\end{equation}
where we used Eq.~\eqref{eq:deterministic} and $x \leq e^x-1$. Thus, the function $f_\text{ss}(\bm{x})$ decreases monotonically along
deterministic trajectories and, since the rate function can always be considered to be bounded by below, it is a Lyapunov function of the deterministic dynamics.

\bibliographystyle{unsrt}
\bibliography{references.bib}

\end{document}